\documentclass{ws-procs975x65}
\usepackage{mathrsfs}

\begin{document}

\markboth{Stefano Ansoldi}%
{Statistical study of emission \emph{versus} activity of AGNs}

\title{A STATISTICAL APPROACH TO THE STUDY\\ OF AGN EMISSION \emph{VERSUS} ACTIVITY\\ (with the detailed analysis of Mrk421)}

\author{STEFANO ANSOLDI}

\address{International Center for Relativistic Astrophysics (ICRA), and\\
Istituto Nazionale di Fisica Nucleare (INFN), Sezione di Trieste, and\\
Dipartimento di Matematica e Informatica, Universit\`{a} degli Studi di Udine,\\
via delle Scienze 206 Udine (UD) I-33100, Italy\\
E-mail: stefano.ansoldi@dimi.uniud.it}

\begin{abstract}
We discuss the theory and implementation of statistically rigorous fits to
synchrotron self Compton models for datasets obtained from multi-wavelength
observations of active galactic nuclei spectral energy distributions. The
methods and techniques that we present are, then, exemplified reporting on a
recent study of a nearby and well observed extragalactic source, Markarian~421.
\end{abstract}

\keywords{Active Galactic Nuclei (AGN); Chi-square minimization; Kolmogorov-Smirnov test; Markarian~421.}

\bodymatter
\vspace*{5mm}
\begin{flushright}
{\small{}Pure logical thinking cannot yield us any knowledge of the empirical world:}\\
{\small{}all knowledge of reality starts from experience and ends in it.}\\
\emph{\small{}Albert Einstein, 1954}.
\end{flushright}
\vspace*{5mm}
\section*{Introduction}

In the last 20 years the \emph{window} on our universe has opened to an unprecedented level,
allowing us to bridge a large fraction of the gap that existed between observational fields,
like astrophysics and cosmology, and more experimental ones, like, for instance, high energy particle
physics. A marked difference between observational and experimental disciplines which is
often emphasized is that the former ones usually lack direct control on the object that we wish
to study, which is usually not directly accessible to us. A consequence of this is that, contrary
to experimental disciplines (where we can mostly prepare the system under study to suit our needs
and tackle the data acquisition and analysis under what we judge to be the most appropriate and
fruitful conditions) in cosmology and astrophysics we usually do not have this convenience. The
lack of control on the systems under study and on their environments can then result in uncertainties
which are usually higher than the ones obtained in experimental situations and make more challenging
to single out discrepancies between models and data. Although this stereotype is true to some
extent, the situation has now radically changed, as witnessed, for instance, by precision
measurements of the cosmic microwave background radiation, or by millisecond pulsar timing.
If it is true that, in general, we have no direct control on the phenomena that take place in our universe,
it has nevertheless to be recognized that they are often probing the laws of physics in such
extreme situations, as we will never be able to realize in a human controlled experiment.
So, while on the theoretical side we are still struggling to get a unified picture of fundamental
interactions at the quantum level, it is undoubted that nowadays this understanding has to also,
if not primarily, face the challenge to be consistent with astrophysical and cosmological data,
in addition to those provided by human-scale experiments: in this sense, a synergic interplay between
large and small scale physics is already a reality.

With this higher perspective in mind, we will more modestly present, in this contribution, an example
of how current observations have the potential to constrain emission models for Active Galactic Nuclei.
Our emphasis will be on the importance of a solid statistical analysis, a precious approach to obtain
a quantitative insight and guide us in challenging and, hopefully, disproving existing models in favor
of more refined and realistic ones.

Our presentation will try to be reasonably self-contained, in the sense that we will cover all the
required topics in what follows: we will, however, not be able to cover them with the required depth,
some of which can be obtained starting from the list of references. We will start in Sec.~\ref{sec:agn},
summarizing some key facts about the sources that we will consider here, namely active galactic nuclei.
Some more details will be given for blazars and their emission models in Subsec.~\ref{sec:blazars}.
We will then continue with a concise presentation of algorithms for (least square) minimization in
Sec.~\ref{sec:min}, to finally introduce a solid standard, which is the Levenberg-Marquardt approach
(Subsec.~\ref{sec:levmar}). The problem of statistical significance is then briefly addressed in
Sec.~\ref{sec:stasig}, where the Kolmogorov-Smirnov test is analyzed (Subsec.~\ref{sec:kolsmi}): other methods to test
how \emph{bad} a fit is\footnote{The commonly used name for these tests is \emph{goodness of fit}
tests, although, technically, the meaningful results are those in which these tests \emph{fail},
showing in this way that the model needs to be refined.} do exist, but will not be discussed here.
Following these prerequisites, Sec.~\ref{sec:Mrk421} presents details of a recent analysis in which
multi-wavelength datasets of the active galactic nucleus Markarian~421 where fitted to a given
emission model: this section contains three subsections, corresponding to each of the topics that are
presented in the three background sections\footnote{Numbering has be chosen so that each subsection
index matches the number of the section in which the corresponding topic is discussed in general.}:
Subsec.~\ref{sec:sou} describes the source and the chosen datasets, Subsec.~\ref{sec:chi} describes
the fit algorithm and the fit results, Subsec.~\ref{sec:sta} describes the statistical significance
of the results.

\section{\label{sec:agn}Active Galactic Nuclei}

Active galactic Nuclei are galaxies characterized by a core that
appears to be brighter and more energetic than that of other galaxies.
It is, indeed, rather common that the core luminosity competes, or even
exceeds, the one of the host galaxy: in some cases it seems that
a luminosity of the order of $10 ^{4}$ times the one of a standard
galaxy can be associated with a region with a linear size significantly
smaller than $1\,$pc. In addition, non-thermal AGNs emission covers a very
broad range of the spectrum.

It is usually believed that the engine at the core of AGNs is a black-hole
with a total mass of millions or billions of solar masses: such an object is
called a supermassive black hole. Although there is only indirect
evidence that this supermassive core (SMC) is a black hole, there is
usually agreement about the fact that the energy necessary to sustain
a luminosity as large as the one mentioned above is coming from the infall of surrounding
matter onto the SMC: this process is generically called accretion. The details of matter
accretion onto the SMC are still to be understood.

Another striking feature of AGNs is, in some cases, the presence of two powerful,
highly collimated jets that shoot out in opposite directions. The way in which
these jets are formed and other fundamental aspects, for instance related to
their composition and to the details of the processes involved in the acceleration
of particles inside these jets, are also subject of active research.

From this short, qualitative, introduction it appears that there is still a lot
that we have to understand about AGNs. Nevertheless, several important
steps forward have been made and we will concentrate, in what follows, on what we
think we do actually understand. From an historical perspective, it is important
to recall that what nowadays we call AGNs comprises a very wide class of extragalactic
objects, which observationally can (and do) appear very different one from the other. For this
reason, we will report below a standard classification scheme\cite{bib:MAGIC}.

There are several characteristics that can be used to classify AGNs, and their common
ground is that they are not usually found in standard galaxies. We already mentioned
one of them, which is the \emph{high luminosity}. AGNs can reach luminosities of the
order of $10^{48}\,\mathrm{erg} \cdot \mathrm{s} ^{-1}$, which is $10,000$ times the
average luminosity of a galaxy (intended as the characteristic luminosity of the field
galaxy distribution). This is, the apparent luminosity, which is what we see after the
radiation emitted has been, for instance, absorbed by the circum-object medium: the
intrinsic luminosity could then be even higher. The energy output is also distinctive,
and characterized by a \emph{broadband continuum emission}. Standard galaxies have a
spectrum which, at zero-th order can be considered the sum of the black body spectra of
all the stars composing the galaxy: since each star spectrum is \emph{approximately} a
black-body spectrum with characteristic temperature that equals the surface temperature
of the star, and since the surface temperature of stars spans a range of about one decade,
then a typical galaxy power output is emitted within about one decade of frequency. On the
contrary, AGNs can have a spectrum ranging from the mid-infrared to the hardest X-rays, and
such that a narrower frequency band dominating the emission is missing. AGN emission can be
moreover characterized by \emph{prominent emission lines}, another element of contrast
with most standard galaxies. The broadness of AGNs spectral lines can be
both, i) very sharp or ii) present broad wings (with variability spanning two orders of
magnitudes, depending on the source). The emission also presents a marked \emph{variability}
at high frequencies (whereas, in the optical band, variability is about 10\% on human life
timescale): there is, nevertheless, a small class of AGNs (which will be mainly interested
in what follows) having a much more marked, i.e. with much shorter timescales, variability.
Although the \emph{polarization} of AGNs emission is weak, it can be statistically appreciated
when compared with that of a standard galaxy: a frequency dependence of polarization is also
usually detected. Finally, AGNs can also be characterized by a strong \emph{radio emission},
a phenomenology that has been extensively studied since the beginning of radio astronomical
observations.

An AGN is a source that presents some (not necessarily all) of the above
properties. For instance, the vast majority of AGNs have a spectrum characterized by a broadband
continuum emission, with strong emission lines, some variability and weak polarization. Many of them
also present a small angular size. Finally, in some cases radio emission has
been detected and, occasionally, the variability and the polarization are both strong.

Several objects fall within the above definitions and we will see
later on that it is possible to devise a tentative unification scheme, the so called \emph{unified
model}. We will come to it after a concise description of the diverse observational phenomenology.

A first kind of objects that shows properties typical of AGNs are the so-called quasi stellar
sources, or \emph{QUASAR}s. QUASARs show two evident relativistic jets, but their main characteristic
is a very luminous and unresolved nucleus with angular size smaller than the arc-second. Radio
emission may be also present, in which case they are called, \emph{radio QUASAR}s. QUASARs without
radio emission are instead called \emph{radio quiet QUASAR}s and they are about $20$ times more
common than radio QUASARs. Both radio and radio quiet QUASARs are found at high redshift and
without signs of a surrounding galaxy.

A second kind of objects in the AGN class are \emph{radio galaxies}. Radio galaxies
are characterized by a radio emission which is thousands times the radio emission of a standard galaxy:
the radio emission is also apparent in two lobes, that extend in opposite directions outside a bright
radio core, to which they appear to be connected by emission jets.
In standard galaxies the radio emission can be traced to the production of relativistic electrons in
supernovae explosions. In radio galaxies, instead, the radio emission, which is characteristically
non-thermal, can be identified with synchrotron emission from ultra-relativistic particles. The spectrum,
which is a broad continuum, is markedly non-thermal also in the optical range, where it especially features
superimposed emission lines.

With some properties similar to those of radio quiet QUASARs, \emph{Seyfert galaxies} are also AGNs, but
having a lower luminosity. A further classification within Seyfert galaxies can be made based on their
spectrum, that can feature i) broad band regions as well as narrow lines (these objects are called Seyfert 1 galaxies)
or ii) only broad lines (these galaxies are named Seyfert 2). Seyfert galaxies are so close to radio quiet
QUASARs that most of the times it is the way in which they are discovered that makes them members of one class
or the other: in particular, in presence of a high redshift and the absence of a surrounding galaxy the objects
tends to be classified as a QUASAR, while broad emission cores in known galaxies are usually classified as Seyfert
galaxies.

Finally we come to the last group of objects in the AGN family, namely \emph{blazars}. Blazars are the most
energetic sources that can be observed in the sky. They are also characterized by two powerful jets
shooting out in opposite directions at relativistic speed: their peculiarity is that we are able
to view one of them at a small angle to the object axis (which is also the jet emission direction), which means
that the jet is practically pointed at us. The emission spectrum is extremely broad, from radio to $\gamma$
frequencies and allows an additional distinction: i) when the spectrum is completely missing emission lines
we have the, so-called, \emph{BL Lacertae} (\emph{BL Lac}) \emph{objects}, which are radio loud objects with marked variability and
strong optical polarization; ii) when, instead, the spectrum also features strong broad emission lines in the
optical range, we have what are called \emph{Optical Violent Variable objects}, which in the radio band look
similar to radio quasars.

We will discuss in more detail some features of AGNs emission, with particular emphasis on blazars, in
subsection \ref{sec:blazars}. Before moving to this topic, we would like, however, remark that
despite their apparent difference form the observational point of view, a unified model to interpret all these
observational features has been proposed; according to this unified model, all the different features of AGNs
are related to the orientation of the source with respect to the observer\cite{bib:1995PASP107803UrryPadovani}.
All AGNs would then be compact supermassive objects and they
would be surrounded by an \emph{accretion disk} composed of hot plasma emitting thermal radiation. The broad
emission lines that are detected in some spectra, mostly in the optical range, would be caused by the presence
of a toroidal shape region containing dense molecular clouds (called the \emph{broad emission lines region}): the
broad emission lines region can obscure the view of the central part. At a larger distance from the central
object there is also a \emph{narrow emission lines region} also filled with molecular clouds but of density lower
than the one present in the broad emission lines region. In a direction perpendicular to the plane of the accretion
disk and in opposite directions, two jest of ultra-relativistic particles emerge from the central object and
extend several kilo-parsecs away from the core. As we anticipated this model allows for explanation of the AGNs
phenomenology, once we consider the orientation of the object with respect to the line of sight of the observer.
Radio galaxies and Seyfert galaxies have the jets nearly orthogonal to the line of sight of the observer: the torus
surrounding the core of the object obscures it and reprocesses the radiation coming from the disk and from the
broad emission lines region. The lobes define the region where the jets loose collimation, and both are responsible
for a strong radio emission. A completely different picture of the same model is obtained if we change the orientation
so that the jets now point in the direction of the observer's line of sight: now the boosted jet emission pointing
directly to the observer is dominating over all the other components, and it is the physics of the jet that is
mostly detected. At intermediate angles between the directions in which the line of sight is perpendicular or parallel
to the jets direction, both the central objects and the surrounding regions are seen: these objects are quasars,
and their characteristic emission lines are the result of the light coming from both the broad and narrow emission
line regions.

We will now provide more details on the emission model, with particular emphasis on blazars.

\subsection{\label{sec:blazars}Blazars and emission models}

Among the realizations of AGNs that we have briefly described above a very important role is played by blazars.
Although only in $\lesssim 10 \%$ of the cases it is possible to detect a jet structure in AGNs, a lot of effort
has been put in understanding the origin of the jets and the physical processes that allow for particle acceleration
in such highly collimated beams. A full modelling of AGNs and AGNs features formation and evolution, is
still one of the fundamental open problems in astrophysics; however, it is currently accepted that jets consist of low entropy
flows associated to regions of internal/external shocks, within which the jets dissipate part of their energy.
Since a relativistic jet that moves in a direction that forms a small angle with the line of sight of the observer
is greatly amplified by \emph{relativistic beaming}, blazars observations are an exceptional tool to investigate
the nature and properties of AGNs relativistic jets: indeed blazars' emission is dominated by the jet and makes
these objects the major extragalactic source of $\gamma$-rays, despite the fact that they represent only a minority
of the AGNs population. Two major subclasses can be identified within blazars of the BL Lac type: while all of them show a spectral
energy distribution (SED) with two pronounced bumps, these bumps peak i) in the infrared the first and around GeV
frequencies the second or ii) in the X-ray the first and around TeV frequencies the second. In particular,
BL Lac objects of the first kind are called \emph{low frequency peaked} BL Lac, while BL Lac objects of the
second kind are called \emph{high frequency peaked} BL Lac (\emph{HBL}).

It is generally agreed that the low energy peak is produced by synchrotron emission. Models can differ, instead,
in the description of the very high energy $\gamma$-ray bump, and can be classified in two qualitative different
groups: \emph{leptonic models} and \emph{hadronic models}.
\begin{description}
    \item[Leptonic models]$\!\!$. These models explain the very high energy $\gamma$-ray radiation by inverse
Compton scattering of photons off relativistic electrons/positrons. If the scattered photons are synchrotron
photons created by the electrons of the jet, the models are called \emph{synchrotron self Compton} (\emph{SSC})
models\cite{bib:1985ApJ298114Marscher,bib:1992ApJL397L5Maraschi,bib:1996ApJ461657Bloom}.
If, instead, the scattered photons are \emph{ambient} photons or photons coming from the environment,
the models are called \emph{external inverse Compton} (\emph{EIC}) models. The absence of emission lines in
the blazars spectra seems to favor SSC models. SSC models can invoke one region in which the relevant interactions
occur (\emph{one-zone} SSC models), or be refined to take into account more than one region. In the following
we are going to test a one zone SSC model on a set of nine simultaneous multi-wavelength SEDs.
    \item[Hadronic models]$\!\!$. In this case the models take advantage from the fact that a proton component
in the jet would be subjected to less synchrotron losses as compared to electrons and could then be accelerated
more efficiently\cite{bib:2005CJAA5195Rieger}. The low energy peak is again explained by synchrotron radiation, so in these models there is
an electron components that is accelerated together with protons. The higher energy bump is instead produced
by the interaction of the accelerated protons with matter and/or ambient photons and or magnetic
fields\cite{bib:1992AAL253L21Mannheim,bib:1993ApJL402L29Bednarek,bib:1993AA26967Mannheim,bib:1997ApJL478L5Dar,bib:1998Sci279684Mannheim,%
bib:2000NAS5377Aharonian,bib:2000AA354395Pohl,bib:2001APP15121Mueche,bib:2003APP18593Mueche}.
Proton induced cascades, that can in turn induce electromagnetic cascades, and/or proton synchrotron models
have also been considered\cite{bib:1992AAL253L21Mannheim,bib:2001APP15121Mueche,bib:2003APP18593Mueche}.
\end{description}
It is also not excluded that both processes could coexist in the jet\cite{bib:2000ApJ534109Sikora,bib:2005ApJ625656Georganopoulos}.
In what follows we will be interested in
a leptonic model, specifically a one zone SSC model\cite{bib:1998ApJ509608Tavecchio,bib:2010MNRAS4011570Tavecchio}.
This model has shown a good agreement with HBL broadband emission in both, the ground and excited
states\cite{bib:2008ApJ6791029Tagliaferri}. Moreover, in one zone SSC models, since there is only one population of electrons
that generates the doubly peaked emission, there is naturally a correlation between the X-ray and the very high energy
$\gamma$-ray variability. The emission zone is assumed to be a spherical blob of radius $R$, moving with a bulk
Lorentz factor $\Gamma$ in a direction forming an angle $\theta$ with respect to the observer viewing direction.
Special relativistic effects can then be described by a single parameter $\delta = (\Gamma (1 - \beta \cos \theta)) ^{-1}$.
The model assumes that the spherical blob region is uniformly filled by electrons with density $n _{\mathrm{e}}$ and
by a uniform, tangled, magnetic field $B$. Five more parameters complete the model providing a description of the relativistic
electrons' spectrum. This is characterized by a smoothed, broken power-law in $\gamma$, the Lorentz factor of the electrons,
which is bounded by $\gamma _{\mathrm{min}} < \gamma _{\mathrm{max}}$. The transition between the two power-laws takes place at $\gamma _{\mathrm{br}}$.
The energy slope at low and high energies are, respectively, $n _{1}$ and $n _{2}$. Altogether the model has nine free parameters:
three of them ($\delta$, $R$ and $B$) describe the emitting blob; the other six ($n _{1}$, $n _{2}$, $\gamma _{\mathrm{min}}$, $\gamma _{\mathrm{max}}$,
$\gamma _{\mathrm{br}}$, $n _{\mathrm{e}}$) describe the energy distribution of the electrons' plasma.

It is crucial to realize that only \emph{simultaneous}, \emph{multi-wavelength} observations allow the determination of all
the parameters of the model. In particular, if we did not have the very high energy $\gamma$-ray observations that became
available with modern Cherenkov telescopes, we would have only knowledge of the synchrotron peak. This would give us
information about the electrons' distribution, but not on the other parameters of the model, and certainly it would not
help us to remove, for example, the residual degeneracy between the intensity of the magnetic field and the electron density.
This simple example already gives an idea of the importance of simultaneous multi-wavelength observations across all the emission
spectrum of the object.

To determine the parameters of the model that we just described, we will use a rigorous statistical approach, by fitting the
SSC model to several simultaneous multi-wavelength datasets corresponding to different activity states of the source. In this
way we will be able to constrain, within some significance level, the SSC model in each emission state (this is the primary
goal of this contribution). In turn, this allows also an analysis of the behavior of the parameters of the model in different
emission states, a point which is discussed in detail elsewhere\cite{bib:2011ApJ73314Mankuzhiyil}.

Before continuing with the analysis anticipated above, we will introduce in the following sections some basic ideas about non-linear fits and
their significance.

\section{\label{sec:min}Nonlinear $\chi ^{2}$ fitting}

In this section we discuss a technique that allows to identify local minima of real valued functions
and that we will use in the context of non-linear least squares problems: in particular, our final goal
will be to use this technique to minimize the sum of the squares
of the deviations between a given set ${\mathcal{O}} = \{ (x _{i}, y _{i} , \sigma _{i}) | i = 1 , \dots , N\}$
of measured data points (in which $\sigma _{i}$ is the uncertainty in the measured quantity
$y _{i}$ and uncertainties in $x _{i}$ are considered small enough to be neglected) and a model
function $f ( x ; {\mathbf{p}})$ that depends from a set
${\mathbf{p}} = ( p _{j} ) _{j = 1 , \dots , n }$ of $n$ parameters. Under suitable assumptions
it can be proved\cite{bib:2003draeaftpsBevington} that the optimal values for the parameters according to the observed data
can be obtained by minimizing the Chi-Square function, i.e.
\[
    \chi ^{2} ( {\mathbf{p}} )
    =
    \frac{1}{2} \sum _{i} ^{1,N}
        \left[
            \frac{y _{i} - f (x _{i} ; {\mathbf{p}})}{\sigma _{i}}
        \right] ^{2}
    .
\]
When the function $f$ is a linear function of the parameters, a closed formula for the minimization of
$\chi ^{2}$ can be obtained. Here we are instead interested in the situation in which $f$ is a non-linear
function of the $p _{j}$. The minimization process can then be performed numerically in several iterations,
the goal of each iteration being to find a perturbation $\delta p _{j}$ of the current values of the parameters
$p _{j}$ that results in a lower value of $\chi ^{2}$. Several methods can be developed to find the values
of the parameters $p _{j}$ that minimize $\chi ^{2}$. In view of our final goal, we will concentrate on two
of these methods, namely the \emph{steepest descent method} and the \emph{Gau\ss{}-Newton} or \emph{Inverse
Hessian} method.

\subsection{The steepest descent method}

The steepest descent method\cite{bib:2004trIMMMadsen,bib:1992nricPress} is based on the evaluation of the gradient
of the objective function (the $\chi ^{2}$
in our case) with respect to the parameters $p _{j}$. The main idea of the method is that
the most direct path in the direction of a local minimum is to descend in the direction opposite to the
gradient of $\chi ^{2}$ with respect to the $p _{j}$. The components of the gradient of $\chi ^{2}$, i.e.
$\partial _{p _{j}} \chi ^{2}$ turn out to be
\[
    \partial _{p _{j}} \chi ^{2}
    =
    - \sum _{i} ^{1,N}
        \left [
            \frac{y _{i} - f (x _{i} ; \mathbf{p})}{\sigma _{i} ^{2}}
            \partial _{p _{j}} f (x _{i} ; {\mathbf{p}})
        \right]
    .
\]
For later convenience we will set up a more compact matrix notation for the above relation, which is
\[
    ( \mathrm{\mathbf{grad} _{{\mathbf{p}}}} \chi ^{2} ({\mathbf{p}}) ) ^{T}
    =
    - ( \mathbf{y} - \mathbf{f}) ^{T} \, \mathbf{\Sigma} \, \mathbf{J}
    ,
\]
where the gradient is nothing but
\[
    \mathrm{\mathbf{grad} _{{\mathbf{p}}}} \chi ^{2} ({\mathbf{p}})
    =
    ( \partial _{p _{j}} \chi ^{2} ) _{j = 1 , \dots , n}
    ,
\]
$\mathbf{y}$ is the $N$-dimensional constant vector of the dependent variable observed data,
\[
    \mathbf{y} = ( y _{i} ) _{i = 1 , \dots , N} ,
\]
$\mathbf{f}$ is the $N$ dimensional vector of the dependent variable values estimated according to the
model described by $f (x , {\mathbf{p}})$,
\[
    \mathbf{f} = \mathbf{f} ({\mathbf{p}}) = ( f (x _{i} ; {\mathbf{p}}) ) _{i = 1 , \dots , N} ,
\]
$\mathbf{\Sigma}$ is the $N \times N$ diagonal matrix of the weights corresponding to the dependent variable
uncertainties in the $N$ measurements ${\mathcal{O}}$,
\[
    \mathbf{\Sigma} = \mathrm{diag}( \sigma _{i} ^{-2}) _{i = 1 , \dots , N}
\]
and, finally\footnote{In this notation the $\chi ^{2}$ function can be written as
\begin{eqnarray}
    \chi ^{2} ({\mathbf{p}}) & = & \frac{1}{2} (\mathbf{y} - \mathbf{f}) ^{T} \, \mathbf{\Sigma} \, (\mathbf{y} - \mathbf{f})
    \nonumber \\
    & = & \frac{1}{2} \mathbf{y} ^{T} \mathbf{\Sigma} \mathbf{y} + \mathbf{y} ^{T} \mathbf{\Sigma} \mathbf{f} +
          \frac{1}{2} \mathbf{f} ^{T} \mathbf{\Sigma} \mathbf{f}
    .
\end{eqnarray}
}, $\mathbf{J}$ is the Jacobean matrix of $\mathbf{f}$, i.e.,
\[
    \mathbf{J}
    =
    ( \partial _{p _{j}} \mathbf{f} ({\mathbf{p}}) ) _{j = 1 , \dots , n}
    =
    ( \partial _{p _{j}} f (x _{i} ; {\mathbf{p}}) ) _{\begin{matrix}{\scriptstyle{}i = 1 , \dots , N} \hfill \\[-2mm] {\scriptstyle{}j = 1 , \dots , n} \hfill \end{matrix}}
    .
\]
A perturbation $\delta \mathbf{p} = (\delta p _{j}) _{j = 1 , \dots , n}$ that updates the parameters in the direction
of the steepest descent, i.e. in the direction opposite to the gradient of $\chi ^{2}$, can then be obtained as
\[
    \delta \mathbf{p} = \mu \mathbf{J} ^{T} \, \mathbf{\Sigma} \, ( \mathbf{y} - \mathbf{f}),
\]
where we used the fact that, since $\mathbf{\Sigma}$ is diagonal, $\mathbf{\Sigma} ^{T} = \mathbf{\Sigma}$.
$\mu$ is a positive real number that determines the length of the step in the steepest descent direction.

Based on the above framework, the steepest descent method consists of a sequence of parameters updates that
are always performed in the direction of the steepest descent until a minimum is found with the prescribed
accuracy. For simple objective functions the steepest descent method is recognized as a highly convergent
approach to minimization and, when the number of parameters is high or very high, it can be considered as
the most reliable, if not the only viable, method. There are, nevertheless, weak points of this method,
especially for complex models: these weak points are substantially related to the fact that the method
does not take into account the curvature of the surface which, in our case, is the graphic of the $\chi ^{2}$. Because of
this, it is possible to make too large steps in steep regions or, too small steps in shallow regions: this
clearly affects the convergence of the algorithm. At the same time, particular structures in the $\chi ^{2}$
surface, as for instance narrow valleys, may also damage convergence. In the case of a narrow valley, for instance,
we would need to move a large step in the direction that points along the flat base of the valley, but only
a small one in the direction perpendicular to the valley walls. If it is true that second order information,
i.e. the use of curvature information about the $\chi ^{2}$ surface, would definitely help to improve the
method, it is often (and as we will see later on, in our case in particular) computationally expensive to
access this second order information. We will see that a good compromise can be found: to this end, we need to
first analyze another approach to minimization, which we will do in the following subsection.

\subsection{The inverse Hessian method}

Another approach that can be used to determine a minimum (in particular of the $\chi ^{2}$ function
described above) is the inverse Hessian (or Gauss--Newton) method\footnote{We will use the first denomination
here, although the second is also quite widespread.}. To give a sound motivation to this
approach\cite{bib:1992nricPress,bib:1996nmflspBjorck}
let us first consider a particular case, i.e. the one in which the dependence of the model function from the parameters
is linear. The model function can then be written as
\[
    f (x , \mathbf{p}) = f ^{(0)} (x) + ( \mathbf{L} ^{(1)} (x) ) ^{T} \mathbf{p}
    \stackrel{\scriptstyle\mathrm{def.}}{=} f ^{(0)} (x) + \sum _{j} ^{1 , n} L ^{(1)} _{j} (x) p _{j}
    ,
\]
where $\mathbf{L} ^{(1)} (x)$ is the vector $\mathbf{L} ^{(1)} (x) = (L ^{(1)} _{j} (x)) _{j = 1 , \dots , n}$.
For this linear model
\[
    \mathbf{f} = \left( f ^{(0)} (x _{i}) + ( \mathbf{L} ^{(1)} (x _{i}) ) ^{T} \mathbf{p} \right) _{i = 1 , \dots , N}
\]
and
\[
    \mathbf{J} = ( (\mathbf{L} ^{(1)} (x _{i}) ) ^{T}) _{i = 1 , \dots , N}
\]
so that
\[
    \mathbf{f}
    =
    \mathbf{f} ^{(0)} + \mathbf{J} \mathbf{p} , \quad
    \mathbf{f} ^{(0)} \stackrel{\scriptstyle\mathrm{def.}}{=}
    (f ^{(0)} (x _{i})) _{i = 1 , \dots , N}
    .
\]
The $\chi ^{2}$ is then a quadratic function of the parameters,
\[
    \chi ^{2} (\mathbf{p}) = \frac{1}{2} ( \mathbf{y} - \mathbf{f} ^{(0)} ) ^{T} \mathbf{\Sigma} ( \mathbf{y} - \mathbf{f} ^{(0)} )
                           - ( \mathbf{y} - \mathbf{f} ^{(0)} ) ^{T} \mathbf{\Sigma} \mathbf{J} \mathbf{p}
                           + \frac{1}{2} \mathbf{p}^{T} \mathbf{J}^{T} \mathbf{\Sigma} \mathbf{J} \mathbf{p}
\]
and the minimum can be obtained in closed form by algebraically solving for $\mathbf{p}$ the linear equation
\[
    -
    ( \mathbf{y} - \mathbf{f} ^{(0)} ) ^{T} \mathbf{\Sigma} \mathbf{J}
    +
     \mathbf{p} ^{T} \mathbf{H}
    =
    0
    ,
\]
which was obtained remembering that, since $\mathbf{\Sigma}$ is diagonal,
$\mathbf{H} \stackrel{\scriptstyle\mathrm{def.}}{=} \mathbf{J}^{T} \mathbf{\Sigma} \mathbf{J}$
is an $n \times n$ symmetric matrix. The final result is the minimum point $\mathbf{p} _{\mathrm{min}}$,
\[
    \mathbf{p} _{\mathrm{min.}}
    =
    \mathbf{H} ^{-1} \mathbf{J} ^{T} \mathbf{\Sigma} ( \mathbf{y} - \mathbf{f} ^{(0)} )
    =
    \mathbf{H} ^{-1} \mathbf{J} ^{T} \mathbf{\Sigma} ( \mathbf{y} - \mathbf{f} ) + \mathbf{p}
    .
\]
In this special case (the model function is linear in the parameters and, thus, the $\chi ^{2}$ is
quadratic in $\mathbf{p}$) we have that i) $\mathbf{H}$ is exactly the \emph{Hessian} of the $\chi ^{2}$,
ii) $\mathbf{H}$ it is a constant
matrix (specifically, constant with respect to $\mathbf{p}$) and iii) it is possible to write in closed
form an exact solution for the minimum.

The inverse Hessian method takes advantage of the above result, by using it to deal with the general case,
in which the model function is generic. The way in which this is obtained is by iterating successive steps:
in each of them a linear approximation of the model function around the current values of the parameters
is used, which results in a quadratic approximation to the $\chi ^{2}$. The exact solution to this linearized
model is given by the equation derived just above and will provide us with a new value of the model parameters;
of course, in the fully non-linear case these new values will unlikely realize the $\chi ^{2}$ minimum, but they
might turn out to be a much better approximation to it. By successive steps convergence to the minimum might
eventually be achieved. The advantage of this method, as opposed to the steepest descent method described in
the previous subsection, is that we are here using information related to the curvature of the $\chi ^{2}$
surface that might allow us to reach more quickly the sought minimum. At the same time, if we are not close
enough to the minimum, the linearized model will probably not be a good enough approximation of the fully
non-linear model. Several steps might be required to arrive close enough to the minimum, where the method
is particularly efficient.

\subsection{\label{sec:levmar}Levenberg and Levenberg-Marquardt methods}

As we have discussed in the previous subsections, both the steepest descent method and the inverse Hessian
method have advantages and disadvantages. The first of the two, is iteratively trying to converge toward the
minimum by updating the parameters as
\[
    \mathbf{p} \longrightarrow \mathbf{p} + \delta \mathbf{p}, \quad
    \delta \mathbf{p} = \mu \mathbf{J} ^{T} \, \mathbf{\Sigma} \, ( \mathbf{y} - \mathbf{f}), \quad
    \mu \in {\mathbb{R}} _{+}
    ;
\]
good convergence could be badly affected in situations in which the shape of the $\chi ^{2}$ surface
presents features that require an estimation of quantities related to the surface curvature, which could
be the case in proximity of the minimum. The second method, was, also iteratively, trying to converge to
the minimum by updating the parameters as
\[
    \mathbf{p} \longrightarrow \mathbf{p} + \delta \mathbf{p}, \quad
    \delta \mathbf{p} = \mathbf{H} ^{-1} \mathbf{J} ^{T} \, \mathbf{\Sigma} \, ( \mathbf{y} - \mathbf{f})
    ;
\]
good convergence is more likely achieved, in this case, close to the minimum, when a linear approximation to
the model function is more appropriate, and the $\chi ^{2}$ surface, locally, could be well approximated by
a paraboloid.

A vantage element, common to both models, is that they only require the calculation of the first derivatives
of the model function (out of which $\mathbf{J}$ is made): no second derivatives appear in these methods, which
would allow to spare computational time. Also, from the qualitative analysis that we have done of the two methods,
they appear to be more effective in complementary situations, the steepest descent being likely efficient far
away from the minimum, while the inverse Hessian could provide better convergence close to it. These considerations
strongly motivate Levenberg proposal\cite{bib:1944QAM2164Levenberg},
which is to iteratively update the parameters according to the following rule:
\[
    \mathbf{p} \longrightarrow \mathbf{p} + \delta \mathbf{p}, \quad
    \delta \mathbf{p} = ( \mathbf{H} + \lambda {\mathbb{I}} ) ^{-1} \mathbf{J} ^{T} \, \mathbf{\Sigma} \, ( \mathbf{y} - \mathbf{f}), \quad
    \lambda \in {\mathbb{R}} _{+},
\]
where ${\mathbb{I}}$ is the $n \times n$ identity matrix. The positive real number $\lambda$ is fixed at a small value at the
beginning of the computation: it is, then, dynamically adjusted by the algorithm according to the estimated distance
from the expected minimum. When the algorithm estimates to be far from the minimum, $\lambda$ is progressively
increased so that the contribution of $\mathbf{H}$ to $\mathbf{H} + \lambda {\mathbb{I}}$ becomes negligible and
the method behaves as the steepest descent one. On the contrary, when the algorithm estimates to be closer to the expected
minimum, the value of the parameter $\lambda$ is progressively reduced, until it will be so close to zero that the
contribution of $\lambda {\mathbb{I}}$ to $\mathbf{H} + \lambda {\mathbb{I}}$ will be negligible and, in all respect,
the algorithm will be following an inverse Hessian approach. The quantity that is computed to decide the decrease/increase
in $\lambda$ is the absolute difference between the value of $\chi ^{2}$ and the value of its quadratic approximation,
with the assumption that this approximation becomes better and better close to the minimum.

Following Levenberg, Marquardt proposed a further refinement of the model\cite{bib:1992nricPress,bib:1963SIAMJAM11431Marquardt},
hence the name of what can nowadays
be considered a robust standard for $\chi ^{2}$ minimization, i.e. the Levenberg-Marquardt method. The proposal
of Marquardt was to use, also during the steps closer to the steepest descent method, part of the information about
the curvature of the $\chi ^{2}$ surface encoded into $\mathbf{H}$ (we remark again that $\mathbf{H}$ is not the
Hessian of the $\chi ^{2}$ but it is the Hessian of the quadratic approximation to the $\chi ^{2}$ which is obtained
by linearizing the model function). The update rule for the Levenberg-Marquardt algorithm is, therefore,
\begin{equation}
    \mathbf{p} \longrightarrow \mathbf{p} + \delta \mathbf{p}, \quad
    \delta \mathbf{p} = ( \mathbf{H} + \lambda \mathrm{diag}(\mathbf{H}) ) ^{-1} \mathbf{J} ^{T} \, \mathbf{\Sigma} \, ( \mathbf{y} - \mathbf{f}), \quad
    \lambda \in {\mathbb{R}} _{+}.
\label{eq:levmarupdalg}
\end{equation}
It is this method that we will apply in the study discussed in section~\ref{sec:Mrk421}.

\section{\label{sec:stasig}Statistical Significance}

In this section we would like to discuss some selected topics about what are usually
called \emph{goodness of fit tests}. In particular we will concentrate our attention
on a particular test, the Kolmogorov-Smirnov (KS)
test\cite{bib:1933GIIA41Kolmogorov,bib:1939MS63Smirnov,bib:1943AMS14305Scheffe,bib:1949ProcBSMSP93Wolfowitz,bib:1951JASA4668Massey}.
The KS test can be used to
decide if a data sample is coming from a population with a specific distribution,
and will be thoroughly discussed in section~\ref{sec:kolsmi}. As an introduction to this more
specific topic, we will first recall the general problem. Then, in the next subsection,
we will describe a standard approach that can be used with binned data. This will give
us the possibility to appreciate some crucial difference (and advantages) of the KS test,
which will be presented in the last part of this section.

\subsection{The general problem.}

Let us consider the case in which we are given two sets of data, say ${\mathcal{O}} ^{(1)}$
and ${\mathcal{O}} ^{(2)}$ : we may be interested in quantifying our certainty about the fact
that the two sets of data are coming from populations having the same distribution function.
To be more precise, let us consider the following statement, i.e. the \emph{null hypothesis}
${\mathscr{N}} _{0}$,
\begin{description}
    \item[${\mathscr{N}} _{0}$]$\!$:$\,\,$\textsc{the two sets of data} ${\mathcal{O}} ^{(1)}$ \textsc{and} ${\mathcal{O}} ^{(2)}$
\textsc{are coming from the same population distribution function}.
\end{description}
We are interested in methods that allow us to \emph{disprove} ${\mathscr{N}} _{0}$
\emph{to a certain level of confidence}; if we can succeed in \emph{disproving}
${\mathscr{N}} _{0}$, then we will conclude that ${\mathcal{O}} ^{(1)}$ and
${\mathcal{O}} ^{(2)}$ are coming from different distributions. We remark that disproving
${\mathscr{N}} _{0}$ \emph{to a certain level of confidence} is as far as we can go from
the statistical perspective and that failure in disproving ${\mathscr{N}} _{0}$ only
shows that at the given \emph{level of confidence} it is \emph{consistent} to consider
the two sets of data as coming from the same distribution. In the general statement
that we are discussing we did not make any assumption about ${\mathcal{O}} ^{(1)}$
and ${\mathcal{O}} ^{(2)}$ that, in general, can be coming from two different unknown
distributions. Later on, we will be interested in a particular case, namely the one in
which one of the two distributions is known. In this case, the null hypothesis will be
\begin{description}
    \item[${\mathscr{N}} ' _{0}$]$\!$:$\,\,$\textsc{the set of data} ${\mathcal{O}} ^{(1)}$
\textsc{is coming from a population distributed as ${\mathcal{D}}$}.
\end{description}

\subsection{The Chi-Square test}

The first approach that we will describe is an accepted standard to solve the above problem
for \emph{binned} data\cite{bib:2003draeaftpsBevington,bib:1992nricPress}.
Let us then consider a binning of the sets of data ${\mathcal{O}} ^{(\alpha)}$, $\alpha = 1,2$,
in $N _{\mathrm{b}}$ bins indexed by a set of integers $i \in I$, such data $n ^{(\alpha)} _{i}$,
$\alpha = 1,2$, is the number of observed points of the ${\mathcal{O}} ^{(\alpha)}$ data falling
in the $i^{\mathrm{th}}$-bin (the binning intervals are the same for both sets of measurements).
We can then construct the following estimator
\begin{equation}
    \chi ^{2} [ {\mathcal{O}} ^{1} , {\mathcal{O}} ^{2} ; N _{\mathrm{b}} ]
    =
    \sum _{i \in I \setminus \bar{I}}
        \frac{( r ^{(1)} n ^{(1)} _{i} - r ^{(2)} n ^{(2)} _{i} ) ^{2}}{n ^{(1)} _{i} + n ^{(2)} _{i}}
    ,
\label{eq:chisqutwodat}
\end{equation}
where $\bar{I} = \{ j \in I \, | \, n ^{(1)} _{j} = n ^{(2)} _{j} = 0\}$ is used to exclude
from the sum bins for which the corresponding term would not be well defined and the $r ^{(\alpha)}$,
$\alpha = 1 , 2$, are defined according to the following relations:
\[
    N ^{(\alpha)} = \sum _{i \in I} n ^{(\alpha)} _{i} = \# {\mathcal{O}} ^{(\alpha)}
    , \quad
    r ^{(\alpha)} = \left( \frac{N ^{{(1)}}}{N ^{(2)}} \right) ^{\alpha - \frac{3}{2}}
    , \quad
    \alpha = 1 , 2 .
\]
To consider the correct $\chi ^{2}$ statistics, we also need the number of degrees of freedom,
$\nu$, that can be associated to the test. This number is $\nu = N _{\mathrm{b}} - N _{\mathrm{c}}$,
where $N _{\mathrm{b}}$, from above, is the number of bins, and $N _{\mathrm{c}}$ is
the number of independent constraints that have been imposed on the sets of data.

A similar test can be applied to the case in which we have only one set of data, say ${\mathcal{O}} ^{(1)}$,
and we would like to disprove if ${\mathcal{O}} ^{(1)}$ is coming from a given distribution ${\mathcal{D}}$.
As before let us consider a binning of ${\mathcal{O}} ^{(1)}$ data in $N _{\mathrm{b}}$ bins; again $n ^{(1)} _{i}$
will be the number of ${\mathcal{O}} ^{(1)}$ data falling in the $i^{\mathrm{th}}$-bin. We will, moreover,
define $\delta _{i}$ as the number of data expected in the $i^{\mathrm{th}}$-bin if the data were distributed according
to ${\mathcal{D}}$ (of course, $\delta _{i}$ does not need to be an integer). If $I$ is the set of integers indexing
the binning and $\bar{I} = \{ j \in I \, | \, n ^{(1)} _{j} = d _{j} = 0\}$,
we can define the following estimator
\begin{equation}
    \chi ^{2} [ {\mathcal{O}} ^{(1)} ; {\mathcal{D}} ; N _{\mathrm{b}} ]
    =
    \sum _{i \in I \setminus \bar{I}}
        \frac{( n ^{(1)} _{i} - \delta _{i} ) ^{2}}{\delta _{i}}
    .
\label{eq:chisquonedat}
\end{equation}
We remark that, in this case, there are some potentially not well-defined terms in the sum, i.e.
terms corresponding to bins in which $\delta _{i} = 0$ and $n ^{(1)} _{i} = 0$. These terms mean
that, according to the distribution ${\mathcal{D}}$, there are no results expected in the given bin,
whereas the observed data \emph{do} have occurrences in the bin: this is a simple case in which \emph{it is disproved}
that the data can be obtained from the distribution ${\mathcal{D}}$. As in the former case the number of
degrees of freedom is needed to have the correct $\chi ^{2}$ statistics. If $N _{\mathrm{p}}$ is the number
of parameters required to know the distribution ${\mathcal{D}}$ that have been determined from the data, and
if the occurrences in each bin that are expected from the model, $\delta _{i}$,  are fixed (and \emph{not}
renormalized to match the total number of observed points) then $\nu = N _{\mathrm{b}} - N _{\mathrm{p}}$.
If, instead, the constraint $\sum _{i \in I} n ^{(1)} _{i} = \sum _{i \in I} \delta _{i}$ is imposed, then
$\nu = N _{\mathrm{b}} - N _{\mathrm{p}} - 1$. Additional independent constraints that should be present,
decrease accordingly the number of degrees of freedom.

Wether we are working in the first framework, with two sets of data, or in the second, with only one set of
data and a comparison distribution, we end up with an estimator
($\chi ^{2} [ {\mathcal{O}} ^{1} , {\mathcal{O}} ^{2} ; N _{\mathrm{b}} ]$ or
$\chi ^{2} [ {\mathcal{O}} ^{(1)} ; {\mathcal{D}} ; N _{\mathrm{b}} ]$ respectively) and an associated number
of degrees of freedom $\nu$. In what follows we will write briefly $\chi ^{2} [ N _{\mathrm{b}} ]$, since our
considerations can be applied in the same way to both cases and we wish to explicitly emphasize the dependence
from the binning that we had to perform. In both definitions of $\chi ^{2} [ N _{\mathrm{b}} ]$ the terms in
the sums are not individually normal. However in the limit in which the number of bins, $N _{\mathrm{b}}$, is
large enough, or the number of events in \emph{each} bin is large enough, it is standard practice to consider
the above defined $\chi ^{2} [ N _{\mathrm{b}} ]$ as the sum of the squares of $\nu$ \emph{normal} random
variables of unit variance and zero mean\footnote{This is the reason why the denominator of~(\ref{eq:chisqutwodat})
is twice the average of $n ^{(1)} _{i}$ and $n ^{(2)} _{i}$: indeed, since the variance of the difference of
two normal variables is the sum of the two individual variances, twice of their average is what is required
to obtain for each term of the sum a random variable with unit variance.}.
The Chi-square probability function $Q (\chi ^{2} | \nu)$, i.e. the probability that the sum of the squares
of $\nu$ random \emph{normal} variables with unit variance and zero mean is greater than $\chi ^{2}$, can be
used with $\chi ^{2} [ N _{\mathrm{b}}]$ to test our null hypothesis. $Q (\chi ^{2} | \nu)$ is defined as
\[
    Q (\chi ^{2} | \nu)
    =
    \frac{1}{\Gamma ( \nu / 2 )}
    \int _{\chi ^{2} / 2} ^{+ \infty}
        e ^{-t} \, t ^{-1 + \nu / 2} \, d t
\]
and it is tabulated for convenience in statistics textbooks.

The Chi-Square method we just recalled works with binned data; although it is always possible to obtain
binned data from continuous data, there is often a great deal of arbitrariness in the binning process
and it is likely that the outcome of the test will depend on the binning (a fact which we already
emphasized in our notation by writing $\chi ^{2} ( N _{\mathrm{b}} )$ above). At the same time the Chi-square
method makes an assumption about the \emph{normality} of the data: although this assumption is at the
background of many statistical results, it might not be always satisfied and it would be desirable to
obtain ways to test the null hypothesis without relying on this assumption. This turns out to be
possible for continuous distribution, and an accepted standard is the Kolmogorov-Smirnov (KS) test, which
is discussed in the next subsection.

\subsection{\label{sec:kolsmi}The Kolmogorov-Smirnov test}

As we anticipated the KS test can be used for continuous distributions. It avoids binning,
it does not assume normality, and it is based on the concept of empirical cumulative distribution function
which we will soon introduce. We will start our analysis by considering the case of ${\mathscr{N}} ' _{0}$,
assuming that the results in ${\mathcal{O}} ^{(1)}$ are the results obtained by sampling $N = N ^{(1)}$
independent identically distributed random variables $X _{i}$, $i = 1 , \dots , N$, distributed according
to some unknown distribution ${\mathcal{P}}$. We will denote by $P (x)$ the cumulative distribution
function associated with ${\mathcal{P}}$, i.e.
$P (x) \stackrel{\scriptstyle\mathrm{def.}}{=} {\mathrm{Prob}} (X _{1} \leq x)$.
An \emph{empirical cumulative distribution function} is a way to count how many of the observed points can
be found below the value $x$, and it is defined as
\[
    P _{N} (x)
    =
    \frac{1}{N}
    \sum _{i} ^{1,N}
    {\mathbb{I}}(X _{i} \leq x)
    ,
\]
where ${\mathbb{I}}({\mathscr{C}}) = 1$ \texttt{if} $\,{\mathscr{C}}$ \texttt{is true}$\,$ and ${\mathbb{I}}({\mathscr{C}}) = 0$
otherwise. $P _{N} (x)$ counts the proportion of ${\mathcal{O}} ^{(1)}$ that can be found below $x$ in steps of $1/N$.
By the law of large numbers it can be seen that the proportion of ${\mathcal{O}} ^{(1)}$ that can be found below $x$
tends to the cumulative distribution function $P (x)$, i.e.
\[
    P _{N} (x) \longrightarrow P (x) \quad \mbox{in probability}.
\]
We will now prove a first result\cite{bib:1948AMS19177Feller,bib:1949JASA20343Doob}.
\paragraph{Proposition.} \emph{If $P(x)$ is continuous then the distribution of}
\begin{equation}
    \sup _{x \in {\mathbb{R}}} | P _{N} (x) - P (x)|
\label{eq:supempcumdisfun}
\end{equation}
\emph{does not depend on $P$}.
\paragraph{Proof.}{$\quad$}\\
{\hrule\vspace*{2mm}\small{}\noindent{}We are interested in the behavior of the distribution of~(\ref{eq:supempcumdisfun}), i.e.
\begin{equation}
    {\mathrm{Prob}} \left( \sup _{x \in {\mathbb{R}}} | P _{N} (x) - P (x)| \leq t \right) .
\label{eq:kolsmipro000}
\end{equation}
Let us define the function $P ^{-1} (z) \stackrel{\scriptstyle\mathrm{def.}}{=} \min \{ x | P (x) \geq z \}$,
where $0 \leq z \leq 1$ because this is the range of $P$, and preliminarily calculate
\begin{equation}
    P _{N} (x)
    =
    P _{N} (P ^{-1} (z))
    =
    \frac{1}{N}
    \sum _{i} ^{1,N}
    {\mathbb{I}}(X _{i} \leq P ^{-1} (z))
    =
    \frac{1}{N}
    \sum _{i} ^{1,N}
    {\mathbb{I}}(P ( X _{i} ) \leq z )
    .
\label{eq:kolsmipro001}
\end{equation}
The above expression, contains $P (X _{i})$, which is a uniform distribution on the interval $[0,1]$,
because the cumulative distribution function $F ( X _{1} )$ is given by
\[
    {\mathrm{Prob}} (P (X _{1}) \leq t)
    =
    {\mathrm{Prob}} (X _{1} \leq P ^{-1} (t))
    =
    P (P ^{-1} (t))
    =
    t
    .
\]
This implies that the random variables $Y _{i} = P (X _{i})$, $i = 1 , \dots , N$, are independent and
uniformly distributed on $[0,1]$, so that we can continue the chain of equalities in~(\ref{eq:kolsmipro001})
to obtain
\begin{equation}
    P _{N} (x)
    =
    P _{N} (P ^{-1} (z))
    =
    \dots
    =
    \frac{1}{N}
    \sum _{i} ^{1,N}
    {\mathbb{I}}(P ( X _{i} ) \leq z )
    =
    \frac{1}{N}
    \sum _{i} ^{1,N}
    {\mathbb{I}}(Y _{i} \leq z )
    ,
\label{eq:kolsmipro002}
\end{equation}
in which the last term is independent from $P$.\\
The proof can now be quickly completed by performing a change of variable in~(\ref{eq:kolsmipro000})
and then using the result in~(\ref{eq:kolsmipro002}); we get
\begin{eqnarray}
    {\mathrm{Prob}} (\sup _{x \in {\mathbb{R}}} | P _{N} (x) - P (x)| \leq t) .
    & = &
    {\mathrm{Prob}} \left( \sup _{0 \leq z \leq 1} | P _{N} (P ^{-1} (z)) - P (P ^{-1} (z))| \leq t \right)
    \nonumber \\
    & = &
    {\mathrm{Prob}} \left( \sup _{0 \leq z \leq 1} \left | \frac{1}{N} \sum _{i} ^{1,N} {\mathbb{I}}(Y _{i} \leq z ) - z \right | \leq t \right)
    \nonumber
\end{eqnarray}
which shows that~(\ref{eq:kolsmipro000}) is independent from P, \emph{Q.E.D.}.\\
\hrule
}
\bigskip
The above results imply that \emph{uniformly over} ${\mathbb{R}}$ we have
\[
    \sup _{x \in {\mathbb{R}}} | P _{N} (x) - P (x)| \longrightarrow 0
\]
(i.e. the largest difference between $P _{N}$ and $P$ goes to zero in probability) and
that the distribution of the above supremum does not depend on the \emph{unknown} distribution
of the sample ${\mathcal{O}} ^{(1)}$, i.e. $P$, whenever $P$ is continuous. The final step that
motivates the KS test follows from the observation that, given $x$, the central limit theorem
implies that $\sqrt{N} ( P _{N} ( x ) - P ( x ))$ converges in distribution to a normal
distribution with zero mean and variance $P (x) (1 - P (x))$ (because this is the variance
of ${\mathbb{I}} (X _{1} \leq x)$).
Moreover, $\sqrt{N} \sup _{x \in {\mathbb{R}}} | P _{N} ( x ) - P ( x ) |$ also converges in
distribution, as shown by the following proposition.
\paragraph{Proposition.} \emph{The cumulative distribution function of}
$\sqrt{N} \sup _{x \in {\mathbb{R}}} | P _{N} ( x ) - P ( x ) |$ is such that
\begin{equation}
    \mathrm{Prob} \left( \sqrt{N} \sup _{x \in {\mathbb{R}}} | P _{N} ( x ) - P ( x ) | \leq t \right)
    \longrightarrow
    P _{\mathrm{KS}} ( t ) ,
\label{eq:proempcumdisfun}
\end{equation}
where $P _{\mathrm{KS}} (t) = 1 - 2 \sum _{k = 1} ^{\infty} (-1) ^{k - 1} \exp (- 2 k ^{2} t)$ is the cumulative
distribution function of the Kolmogorov-Smirnov distribution.\\
\paragraph{Proof.}{$\quad$}\\
{\hrule\vspace*{2mm}\small{}\noindent{}The proof of this result will not be given here\cite{bib:1948AMS19177Feller,bib:1949JASA20343Doob}.\\
\hrule
}
\bigskip
The net result of the above analysis is that if $P _{0}$ is the cumulative distribution function
of the distribution associated with ${\mathscr{N}} ' _{0}$, then we can consider the statistics
\begin{equation}
    D _{N} = \sqrt{N} \sup _{x \in {\mathbb{R}}} | P _{N} (x) - P _{0} (x)| ,
\label{eq:kolsmista}
\end{equation}
which will depend only on $N$ and can then be tabulated\cite{bib:1933GIIA41Kolmogorov,bib:1939MS63Smirnov}. If $N$ is big enough the distribution
of $D _{N}$ is approximated by the Kolmogorov-Smirnov distribution. We will now consider what
happens if ${\mathscr{N}} ' _{0}$ fails, which means that $P \neq P _{0}$: in this case the
empirical cumulative distribution function will converge to $P$, it will then not approximate
$P _{0}$ and for large $N$ we will have $\sup _{x \in {\mathbb{R}}} | P _{N} (x) - P _{0} (x)| > \epsilon$
for some small enough, positive $\epsilon$, i.e.
\[
    D _{N} = \sqrt{N} \sup _{x \in {\mathbb{R}}} | P _{N} (x) - P _{0} (x)| > \sqrt{N} \epsilon .
\]
This will allow to define a decision rule in the form $D _{N} \leq c$, where the constant $c$ is
defined by the significance level and the decision rule can be verified by tabulated values for $D _{N}$.

The Kolmogorov-Smirnov test can also be applied to ${\mathscr{N}} _{0}$, where we are interested
in understanding if the two sets of data ${\mathcal{O}} ^{(1)}$ and ${\mathcal{O}} ^{(2)}$ are coming
from the same distribution. Let $\{ X ^{(\alpha)} _{i} \} _{i = 1 , \dots , N ^{\alpha}}$
be the sample of ${\mathcal{O}} ^{(\alpha)}$ having cumulative distribution function $P ^{(\alpha)}$,
$\alpha = 1, 2$. If $P ^{(\alpha)} _{N ^{(\alpha)}}$, $\alpha = 1, 2$, are the corresponding empirical
cumulative distribution functions, then the two propositions above are also satisfied by the following
statistics,
\[
    D _{N ^{(1)} N ^{(2)}}
    =
    \left(
        \frac{N ^{(1)} N ^{(2)}}{N ^{(1)} + N ^{(2)}}
    \right) ^{1/2}
    \sup _{x \in {\mathbb{R}}}
        \left |
            P ^{(1)} _{N ^{(1)}} - P ^{(2)} _{N ^{(2)}}
        \right |
    ,
\]
to which a similar decision rule as the above can be applied.

\subsection{\label{sec:stasigfit}Statistical Significance of $\chi ^{2}$ fits}

In the following we will be interested in determining the statistical significance of fitting
observed data points to a highly non-linear model function. In addition, the model function
will depend from several parameters, and, although at the present level of knowledge these
parameters are considered independent, the complexity of the physical situation makes it
possible that, in more refined models, some of them might show correlations. In addition to the
complex structure of the models, there is the fact that the data points will be spread across
several decades (range of the independent variable) and will come from
different instruments based, not only on different hardware/software, but also on
different physical processes for their operation. It is important under these circumstances
to have a check on the goodness of fit. Because of the complexity of the situation, the approach
that we propose is to perform a Kolmogorov-Smirnov test on the normality of the residuals
obtained after the fitting procedure. In this case our null hypothesis will be
\begin{description}
    \item[\label{text:nulhyp002}${\mathscr{N}} '' _{0}$]$\!$:$\,\,$\textsc{the residuals of the non-linear fit of
multi-wavelength data to the spectral energy distribution function obtained by implementing
a given synchrotron self Compton model are normally distributed}.
\end{description}

As a final remark, we remember that there are situations in which the critical values of the
test statistics can be difficult to calculate: these include situations in which the samples
are of small size and/or the parameters of the distribution are estimated with the same data
that are being tested. For the second case a convenient solution is the inclusion of a correction
factor, but, in general, the safest way to procede is to use Monte Carlo \label{text:MonCar}
methods, for instance to generate, under the fitted distribution, datasets of the same length as the tested one, and
use these them to obtain the correct critical values.

\section{\label{sec:Mrk421}An application: Markarian~421}

As a \emph{field test} application of what we have seen in the previous sections, we will now apply the models and
techniques introduced above to a concrete case, i.e. the study of the emission properties of the AGN Markarian~421,
following a recent publication\cite{bib:2011ApJ73314Mankuzhiyil}. This section will be divided in subsections. In
each of them we will discuss one of the aspects that we have introduced in the sections above and there will be
a direct correlation with the subsection number and the section number in which the concepts exemplified in each
subsection have been discussed in general.

\subsection{\label{sec:sou}The source}

Markarian~421 (Mrk421) is the closest blazar (at a redshift $z = 0.030$) and the first extragalactic $\gamma$-ray source
with emission in the TeV range detected by Imaging Air Cherenkov Telescopes\cite{bib:1992Nature358477Punch,bib:1996A&A311L13Petry}.
It is, nowadays, the most well known blazar, together with the, also close one, Markarian~501, and falls within the class of
HBL objects. Mrk~421 is a source that shows remarkable variability, both in flux variations
(that were observed to change by almost two orders of magnitude) and time development (flux doubling was detected on a time
scale of the order of 15 minutes\cite{bib:1996Nature383319Gaidos}).

For our purpose Mrk421 is an excellent source, which has been extensively studied across over 19 decades in energy. In particular:
\begin{enumerate}
    \item the SSC emission dominates the detected spectrum, with correlated low-high energy fluctuations;
    \item the Compton peak is in the range where Cerenkov observations are effective;
    \item the spectrum can be described as a single power-law.
\end{enumerate}
For the above reasons, the one-zone SSC model that we have described in the previous section is a good candidate to describe
the SED of this source. At the same time simultaneous multi-wavelength observations are available, and, among them, it was
possible to identify nine, good to very good quality, spectral energy distribution sets of simultaneous data corresponding
to different emissions states.
The detailed description of the datasets is reported below.
\begin{description}
    \item[state 1 and state 2]$\!\!$. The first two datasets that we consider correspond to multi-wavelength data obtained from
campaigns triggered by a major outburst of Mrk421 that was detected by the $10$m \emph{Whipple} telescope in April
2006\cite{bib:2009ApJ703169Acciari}. It was unfortunately
impossible to promptly set-up a multi-wavelength campaign because of some visibility constraints on \emph{XMM-Newton}; for this reason
simultaneous observations at different wavelengths were taken during the decaying phase of the flare. The optical monitor of XMM-Newton
was used to collect optical and ultraviolet data, whereas the EPIC-pn detector of the same telescope provided X-ray observations. Very
high energy $\gamma$-ray data were collected by \emph{MAGIC} and Whipple telescopes. Altogether this resulted in more than 7 hours of simultaneous
observations, about 4 hours of which form the datasets for state 1 and more than 3 hours the dataset for state 2.
    \item[state 3]$\!\!$. A third dataset contains multi-wavelength observations initiated after a detection by the all-sky monitor of the
\emph{Rossi X-ray Timing Explorer} and by the $10$m Whipple telescope\cite{bib:2006ApJ641740Rebillot}.
The observation campaign continued during December 2002 and January 2003.
The very high energy data was obtained by Whipple between December 4, 2002 and January 15, 2003 and by the \emph{High Energy Gamma Ray Astronomy
CT1} between November 3, 2002 and December 12, 2003. However, since our analysis is centered on simultaneous observations, we have
used only the very high energy data taken at nights during which simultaneous X-ray observations were available. Optical information
consists of the average flux obtained from the \emph{Boltwood observatory} optical, \emph{KVA} and \emph{WIYN} telescopes.
    \item[state 4 and state 9]$\!\!$. Two more datasets are the result of a longer campaign undertaken during 2003 and
2004\cite{bib:2005ApJ630130Blazejowski}. The
Rossi X-ray Timing Explorer was used to collect the X-ray flux, that was then classified into three sets, having low-, medium- and
high-flux, respectively. X-ray observations where then complemented by Whipple very high energy $\gamma$-ray data, taken within
one hour of the selected X-ray ones. Whipple Observatory $1.2$m telescope and Boltwood Observatory $0.4$m telescope provided optical
data following the same grouping method: although it was not possible to get optical data simultaneously with the remaining multi-wavelength
data, the fact that minor variations in the optical flux were detected, allowed to consider its maximum and minimum values as
reliable approximations. In this way it was possible to obtain a medium flux dataset, corresponding to state 4, between March 8
and May 3, 2003 and a high flux dataset, corresponding to state 9, between April 16 and 20, 2004.
    \item[state 5 and state 7]$\!\!$. Another two datasets are the result of a multi-wavelength campaign that took place between March 18 and
March 25, 2001\cite{bib:2008ApJ677906Fossati}.
X-ray data are the results of Rossi X-ray Timing Explorer observations, whereas the very high energy $\gamma$-ray flux was
obtained by the Whipple telescope. State 5 corresponds to a post-flare state during March 22 and 23, for which optical information
corresponds to the lowest flux detected by the $1.2$m Harvard-Smithsonian telescope on Mt. Hopkins. State 7 is, instead, the high-flux peak
of March 19, also complemented by optical data from the same instrument as for state 5, but using the highest optical flux.
    \item[state 6]$\!\!$. This dataset were taken after an outburst in May 2008 and contains the results of about $2 1/2$ hours of
simultaneous observations\cite{bib:2009ApJ703169Acciari}. As for state 1 and state 2 the optical monitor of XMM-Newton was used to
collect optical and ultraviolet data, whereas the EPIC-pn detector of the same telescope provided X-ray observations. The very high
energy $\gamma$-ray flux was in this case obtained by \emph{VERITAS}.
    \item[state 8]$\!\!$. The last dataset is the result of a dedicated multi-wavelength campaign on June 6, 2008\cite{bib:2009ApJ691L13Donnarumma}.
Optical data was obtained by \emph{WEBT}, whereas X-ray observations were made by the Rossi X-ray Timing Explorer and by \emph{Swift}/BAT and,
finally, very high energy $\gamma$-ray fluxes were taken by VERITAS.
\end{description}
All sets of data present marked qualitative difference between the optical to X-ray and the very high energy $\gamma$-ray ranges:
the most striking one is the uncertainties in the measured flux, which is very small when not negligible at low energies,
and much sizeable at very high energies. This observation will play a role in our future discussion on the statistical significance
of the fit results.
In what follows we will, first, show how to fit each of these datasets to the chosen SSC model and how to test the goodness of the fit.
An in depth interpretation of the physical conclusions about the source is beyond the scope of this lecture and can be found in the
literature\cite{bib:2011ApJ73314Mankuzhiyil}.

\subsection{\label{sec:chi}Nonlinear $\chi^{2}$ fit}

The fit algorithm will be an implementation of the Levenberg-Marquardt method discussed in
subsection \ref{sec:levmar}. As this is a standard approach in nonlinear minimization, it is
possible to conveniently find code which is optimized and efficient for general problems.
In particular we have used as a starting point the \texttt{mrqmin} function discussed in
Ref.~\refcite{bib:1992nricPress}. This subroutines executes a single minimization step,
which is an update step on the parameters as defined in (\ref{eq:levmarupdalg}).
\begin{figure}
\begin{center}
\fbox{\includegraphics[width=12cm]{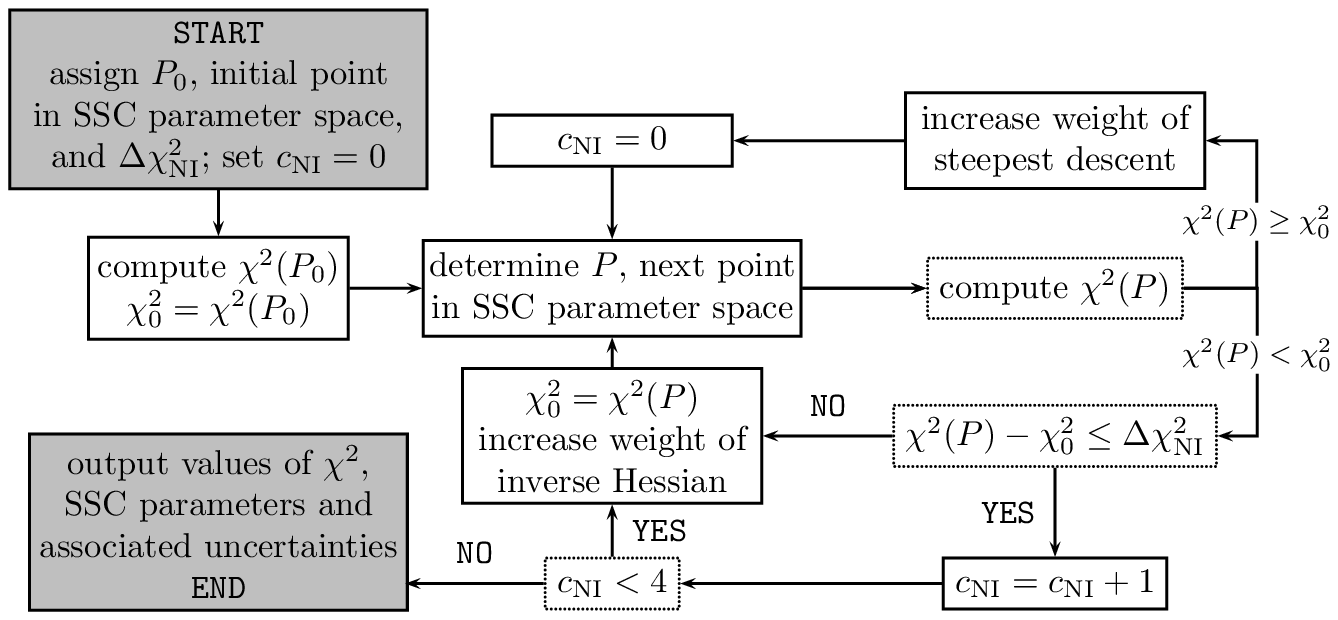}}
\caption{\label{fig:minflocha} Flow chart of the minimization algorithm
(from Ref.~\protect\refcite{bib:2011ApJ73314Mankuzhiyil}). A careful choice of $\Delta \chi ^{2} _{\mathrm{NI}}$
and of the exit value for $c _{\mathrm{NI}}$ is necessary to obtain consistent results avoiding, at the same time,
unnecessary computational cost. The optimal value for the exit condition on $c _{\mathrm{NI}}$ is used in the
flow-chart. $\Delta \chi ^{2} _{\mathrm{NI}}$ as small as $10 ^{-4}$ was required, especially for sets containing
larger number of data points.}
\end{center}
\end{figure}
Implementation details and additional functions called by \texttt{mrqmin} are described in
in Ref.~\refcite{bib:1992nricPress} and will not be discussed here, except for the case of
the code that is used to evaluate the SED, which we will analyze in more detail in a moment.
The single minimization step performed by \texttt{mrqmin} (or by any other implementation
of the Levenberg-Marquardt method) has to be iterated until convergence to the sought minimum
is considered satisfactory. The flow-chart of the code that we used is reported in
Fig.~\ref{fig:minflocha}. After fixing some trial values for the parameters of the model ($P _{0}$
point in parameter space) and initializing to zero an integer variable, $c _{\mathrm{NI}}$ that will
count how many consecutive individual minimization steps have been performed with a negligible
improvement in decreasing the $\chi ^{2}$, we calculate $\chi ^{2} _{0}$, i.e. the value of $\chi ^{2}$
at the initial point $P _{0}$ in parameter space. Minimization iterations then start: at the beginning
we fix the parameter corresponding to $\lambda$ in (\ref{eq:levmarupdalg}) so that we are performing
a step in which the steepest descent contribution to the algorithm is dominant. The new value of
$\chi ^{2}$ at the new point in parameter space $P$ determined by the algorithm, $\chi ^{2} (P)$,
is calculated. If the $\chi ^{2}$ did decrease in the step, we check if it decreased by a sizeable
amount. If not we increment by one the $c _{\mathrm{NI}}$ counter, before increasing the weight
of the inverse Hessian method and moving to the next iteration. If, instead, the $\chi ^{2}$ increased
at $P$, we increase the importance of the steepest descent method and reset the counter $c _{\mathrm{NI}}$
to zero. The exit condition from the loop is satisfied when $c _{\mathrm{NI}} = 4$. Another crucial
parameter that has to be fixed is the threshold, below which we consider negligible the decrease in
$\chi ^{2}$ (this is called $\Delta \chi ^{2} _{\mathrm{NI}}$ in the flow-chart). It is usually
considered that it is unnecessary to determine to machine accuracy or to roundoff limit the
minimum of the $\chi ^{2}$, as the result provides only a statistical estimation of the parameters
anyway. However, in our experience, it is important to pay particular attention in setting the
exit condition value of $c _{\mathrm{NI}}$ and the negligible $\chi ^{2}$ increment, $\Delta \chi ^{2} _{\mathrm{NI}}$.
\begin{figure}
\begin{center}
\fbox{\includegraphics[width=12cm]{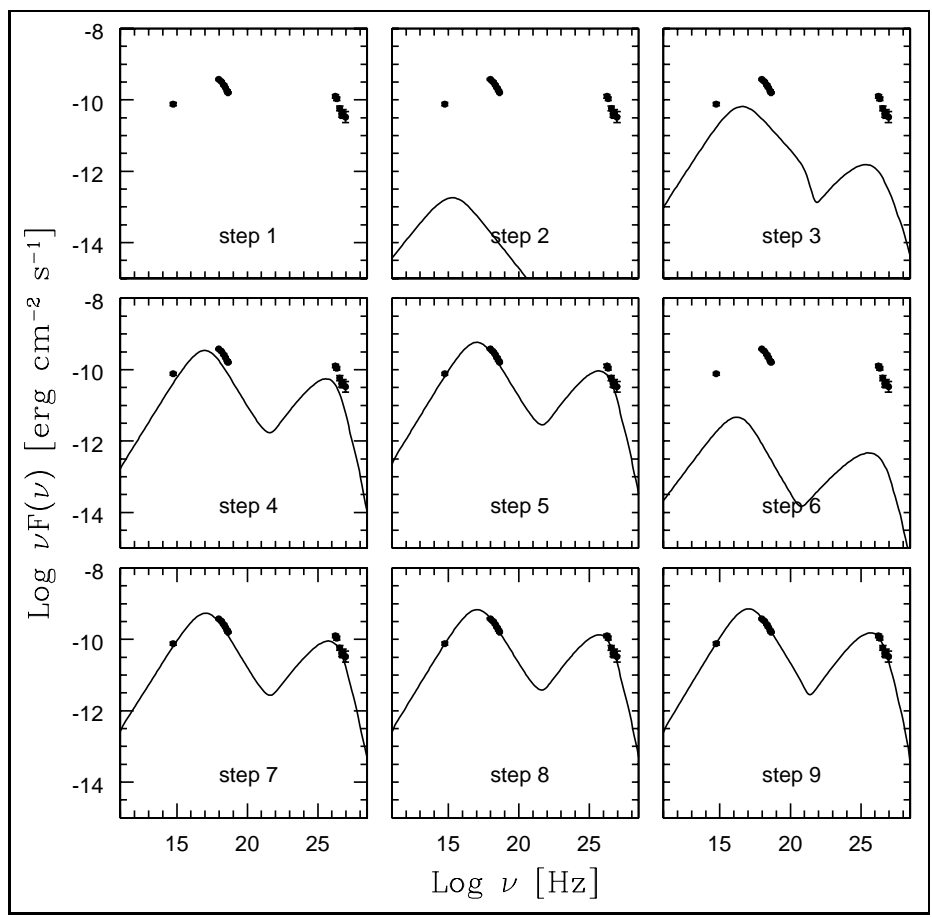}}
\caption{\label{fig:minste}Snapshots of different steps in one run of the minimization code. The
step numbering is conventional and there is no rigorous correspondence between algorithm iteration
number and step number in the figure above, except that a higher step number corresponds to an iteration
that follows the iterations associated to lower-numbered steps (from Ref.~\protect\refcite{bib:2011ApJ73314Mankuzhiyil}).}
\end{center}
\end{figure}
Our experience has shown that the best results were obtained with\footnote{The values of the two
parameters may be correlated, so other choices may work as well.} $c _{\mathrm{NI}} = 4$ and
$\Delta \chi ^{2} _{\mathrm{NI}} = 10 ^{-4}$. As a further test of the minimization algorithm, we
implemented an additional step: after obtaining the minimum we perform a random change of the parameters
and repeat the minimization from this new random point in parameter space. This could help to identify
cases in which minimization remained stuck in a local minimum different from the absolute one, but we never
faced this situation, i.e. the additional minimization step did always converge to the result of the first
one. When we choose smaller/larger values for $c _{\mathrm{NI}}/\Delta \chi ^{2} _{\mathrm{NI}}$ (for instance,
$2$ and $0.01$) the minimization occasionally provided a result which, on closer inspection, turned out to be
a not good approximation to the $\chi ^{2}$ minimum\footnote{Meaning that repeating the minimization with
the optimal values for the parameters resulted in an appreciable different and lower $\chi ^{2}$ (and in
more consistent values for the obtained uncertainties on the parameters, a point which we will discuss in more
detail later on).}. This tendency was extremely more marked in presence of datasets having a larger or much larger
number of data points: in our experience convergence is usually slower in these cases: a tentative visual
representation of some steps in the minimization process is given in Fig.~\ref{fig:minste} (for details, please see
the related caption).

\begin{figure}
\begin{center}
\fbox{\includegraphics[width=12cm]{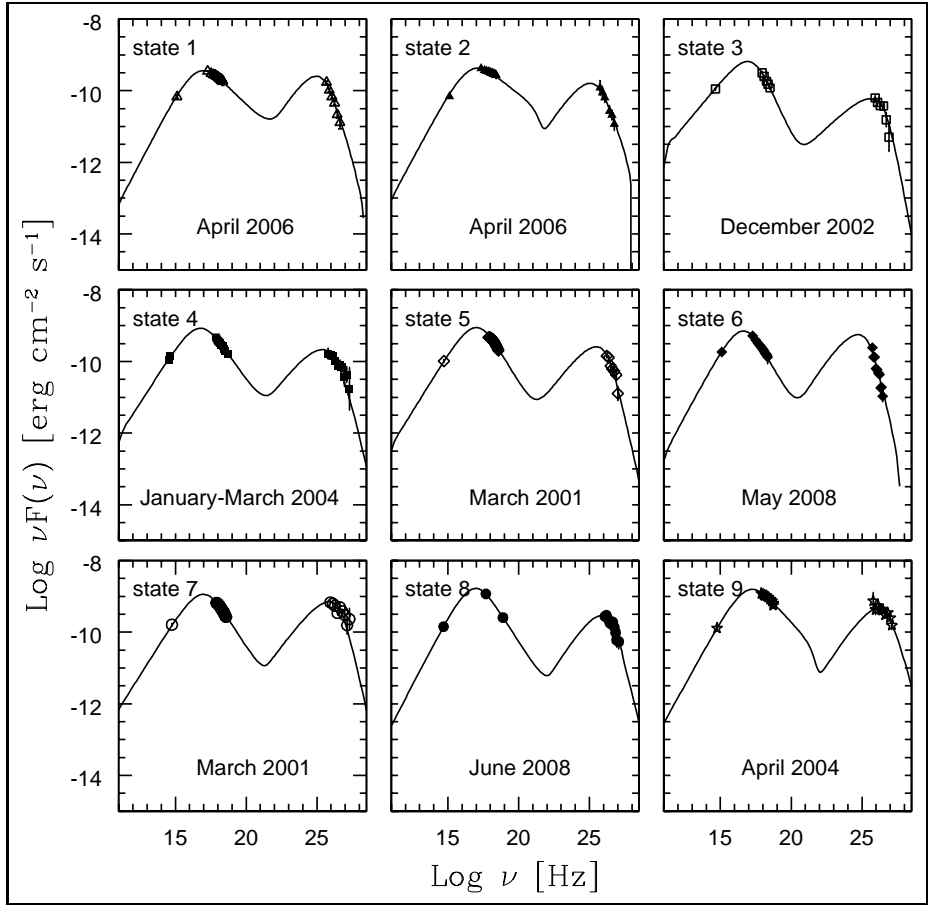}}
\caption{\label{fig:fitsed} Plot of the SEDs obtained for each state with the $\chi ^{2}$ minimization
procedure described in the text (from Ref.~\protect\refcite{bib:2011ApJ73314Mankuzhiyil}). Results for
the parameters and associated uncertainties, together with the values of the reduced $\chi ^{2}$ in
each case, are reported in Tables~{\protect\ref{tab:results01}} and~{\protect\ref{tab:results02}}.}
\end{center}
\end{figure}
From the point of view of the computation time the Levenberg-Marquardt algorithm is quite efficient\footnote{As an
example, the longest minimizations required about $10 ^{2}$ iterations that on a less than average PC (with a
Intel\textregistered{} Core\texttrademark{}2 Duo CPU (E7500, $2.93$GHz) running a x$86$\_$64$ GNU/Linux with $2$GB
Ram) took about 20 minutes to run.} and most of the time was required by the derivation of the SEDs corresponding
to the current values of the parameters at each iteration. In our analysis, following a common assumption in
the literature, we have fixed $\gamma _{\mathrm{min}} = 1$; the redshift of the source is known, so high energy
data points can be corrected to take into account interaction with the extragalactic background light. We are,
then, left with eight parameters to be determined by the fit. According to (\ref{eq:levmarupdalg}) at
each step we need:
\begin{description}
    \item[$\mathbf{y}$]$\!\!$: this is just the vector of the observed flux values, which is clearly known;
    \item[$\mathbf{\Sigma}$]$\!\!$: this is the vector of the flux uncertainties, also known;
    \item[$\mathbf{f}$]$\!\!$: this is the vector of the values of the model SED evaluated at the observed energy/frequency;
    \item[$\mathbf{J}$]$\!\!$: this is a matrix which is known once the derivatives of the model SED are known;
    \item[$\mathbf{H}$]$\!\!$: this matrix can be calculated knowing $\mathbf{\Sigma}$ and $\mathbf{J}$.
\end{description}
It is clear that $\mathbf{y}$, $\mathbf{\Sigma}$ and $\mathbf{H}$ do not represent a problem. $\mathbf{f}$ and
$\mathbf{J}$ in standard cases, where the model function has a known analytical expression, also do not. In our
case (and in several others), however, the model SED is not analytically known and what we have is just a
discretized sample resulting from a numerical implementation of the SSC model; it is this last numerical
implementation that can be more computationally intensive. This is especially true since the estimation of
$\mathbf{J}$ requires the SEDs partial derivatives with respect to the parameters. Then, for each minimization
iteration, we have in principle to evaluate a number of SEDs equal to twice the number of varying parameters
plus one. In our implementation we tried to reduce the load caused by this task by developing a bookkeeping
mechanism that caches some of the numerically sampled SEDs used in the previous steps: this is especially
useful when the iteration results in a $\chi ^{2}$ increase, since in this case, when returning to the previous
point in parameter space, the required SEDs are already available. \emph{Caching} optimizes the number of times at which
the $\chi ^{2}$ minimization executable needs to stop and wait for the completion of an external module that, independently,
executes the SEDs evaluation.
\begin{table}
\tbl{Best-fit single-zone SSC model parameters for the nine datasets of Mrk~421.
States are labelled according to the convention in Fig.~{\protect\ref{fig:fitsed}}
(from Ref.~\protect\refcite{bib:2011ApJ73314Mankuzhiyil}). Here parameters describing
the emitting blob together with the reduced $\chi ^{2}$ are reported. Results for the remaining
parameters can be found in Table~{\protect\ref{tab:results02}}.}
{%
\begin{tabular}{ l c c c | c}
\noalign{\smallskip}
\toprule
\noalign{\smallskip}
Source & $B$ & $R$ & $\delta$ & $\tilde{\chi} ^{2}$ \\
       &[gauss]&[cm]&         & \\
\noalign{\smallskip}
\hline
\noalign{\smallskip}
State 1 & $(9 \pm 3) \times 10 ^{-1}$ & $(9 \pm 4) \times 10 ^{14}$ & $(2.0 \pm 0.5) \times 10 ^{1}$ & $0.84$\\
State 2 & $(8 \pm 6) \times 10 ^{-1}$ & $(8 \pm 4) \times 10 ^{14}$ & $(2.7 \pm 1.1) \times 10 ^{1}$ & $1.86$\\
State 3 & $(6 \pm 6) \times 10 ^{-2}$ & $(2.0 \pm 1.5) \times 10 ^{15}$ & $(1.0 \pm 0.5) \times 10 ^{2}$ & $0.91$\\
State 4 & $(1.21 \pm 0.16) \times 10 ^{-1}$ & $(1.1 \pm 1.3) \times 10 ^{15}$ & $(8 \pm 6) \times 10 ^{1}$ & $0.89$\\
State 5 & $(1.9 \pm 1.3) \times 10 ^{-1}$ & $(10 \pm 4) \times 10 ^{14}$ & $(7 \pm 5) \times 10 ^{1}$ & $0.67$\\
State 6 & $1.0 \pm 0.7$ & $(6 \pm 3) \times 10 ^{14}$ & $(2.8 \pm 1.1) \times 10 ^{1}$ & $1.39$\\
State 7 & $(4 \pm 3) \times 10 ^{-2}$ & $(2 \pm 5) \times 10 ^{15}$ & $(8 \pm 7) \times 10 ^{1}$ & $1.61$\\
State 8 & $(6 \pm 3) \times 10 ^{-2}$ & $(2.0 \pm 1.8) \times 10 ^{15}$ & $(1.1 \pm 0.4) \times 10 ^{2}$ & $0.60$\\
State 9 & $(4 \pm 3) \times 10 ^{-2}$ & $(2 \pm 4) \times 10 ^{15}$ & $(1.2 \pm 1.0) \times 10 ^{2}$ & $0.85$\\
\noalign{\smallskip}
\botrule
\end{tabular}}
\label{tab:results01}
\end{table}

\begin{table}
\tbl{Best-fit single-zone SSC model parameters for the nine datasets of Mrk~421.
States are labelled according to the convention in Fig.~{\protect\ref{fig:fitsed}}
(from Ref.~\protect\refcite{bib:2011ApJ73314Mankuzhiyil}). This table lists the
parameters that describe the energy distribution of the electron plasma. The
parameters describing the emitting blob, together with the reduced $\chi ^{2}$
obtained in the fit, can be found in Table~{\protect\ref{tab:results01}}.}
{%
\begin{tabular}{ l c c c c c }
\noalign{\smallskip}
\toprule
\noalign{\smallskip}
Source & $n_{\rm e}$ & $\gamma_{\rm br}$ & $\gamma_{\rm max}$ & n$_1$ & n$_2$ \\
       & [cm$^{-3}$] &                    &                   &                    &     \\
\noalign{\smallskip}
\hline
\noalign{\smallskip}
State 1 & $(1.3 \pm 1.5) \times 10 ^{3}$ & $(2.6 \pm 0.9) \times 10 ^{4}$ & $(1.05 \pm 0.18) \times 10 ^{7}$ & $1.49 \pm 0.19$ & $3.77 \pm 0.11$\\
State 2 & $(1 \pm 3) \times 10 ^{3}$ & $(2.4 \pm 0.9) \times 10 ^{4}$ & $(4.1 \pm 1.1) \times 10 ^{6}$ & $1.5 \pm 0.3$ & $3.62 \pm 0.14$\\
State 3 & $(5 \pm 5) \times 10 ^{3}$ & $(7 \pm 3) \times 10 ^{4}$ & $(7 \pm 5) \times 10 ^{7}$ & $2.05 \pm 0.10$ & $4.8 \pm 0.3$\\
State 4 & $(2 \pm 5) \times 10 ^{3}$ & $(4 \pm 2) \times 10 ^{4}$ & $(8.2 \pm 1.7) \times 10 ^{6}$ & $1.8 \pm 0.3$ & $4.11 \pm 0.13$\\
State 5 & $(2 \pm 5) \times 10 ^{3}$ & $(4.5 \pm 1.9) \times 10 ^{4}$ & $(2.4 \pm 0.3) \times 10 ^{7}$ & $1.7 \pm 0.3$ & $4.3 \pm 0.18$\\
State 6 & $(4 \pm 4) \times 10 ^{3}$ & $(1.9 \pm 0.6) \times 10 ^{4}$ & $(1.8 \pm 0.4) \times 10 ^{6}$ & $1.54 \pm 0.11$ & $4.37 \pm 0.09$\\
State 7 & $(1 \pm 7) \times 10 ^{3}$ & $(8 \pm 6) \times 10 ^{4}$ & $(7 \pm 2) \times 10 ^{6}$ & $1.7 \pm 0.4$ & $4.23 \pm 0.20$\\
State 8 & $(4 \pm 9) \times 10 ^{1}$ & $(5 \pm 2) \times 10 ^{4}$ & $(1.6 \pm 0.4) \times 10 ^{7}$ & $1.5 \pm 0.2$ & $4.22 \pm 0.14$\\
State 9 & $(1 \pm 7) \times 10 ^{2}$ & $(8 \pm 9) \times 10 ^{4}$ & $(1.1 \pm 0.4) \times 10 ^{7}$ & $1.6 \pm 0.5$ & $3.9 \pm 0.2$\\
\noalign{\smallskip}
\botrule
\end{tabular}}
\label{tab:results02}
\end{table}
A first study obtained applying the $\chi ^{2}$ minimization algorithm on the nine datasets described in
Subsec.~\ref{sec:sou} results in the SEDs plotted in Fig.~\ref{fig:fitsed} and in the values of the
parameters and related uncertainties listed in Tables~\ref{tab:results01} and~\ref{tab:results02}. We
are not interested here in a detailed report of the physical conclusions that can be drawn from these
results, for which we refer the reader to Ref.~\refcite{bib:2011ApJ73314Mankuzhiyil}. We will, instead,
consider some elements relevant to the statistical analysis.

First, reduced $\chi ^{2}$ values are reasonable: a couple of cases (states 5 and 8) might require
some additional check, as values are slightly low. Uncertainties in most cases allow to constrain
parameters within physically meaningful and expected ranges. There are some exceptions, in which uncertainties
tend to be larger than usually acceptable, like in the case of the magnetic field of state 3 or
the blob radius of state 5 and, for most states, the blob electron density. In the cases in which the
uncertainties appear to be too high, it is important to remember that these uncertainties are
obtained by considering a quadratic approximation to the $\chi ^{2}$ near the estimated minimum.
This is a good approximation only when the uncertainties are relatively small, because we can
not expect the $\chi ^{2}$ surface to behave as the quadratic approximation far from the minimum.
It might also happen that, because of the nature of the problem/model, the $\chi ^{2}$ has a much
more flat minimum in the direction of some of the parameters. In all these cases the quadratic
approximation might overestimate the uncertainties and it would be preferable to use the criterion
that gives the uncertainties as the absolute difference between the minimum value of each parameter
at the estimated minimum and the value of
the parameter at which the $\chi ^{2}$ has increased by one. This definition of the uncertainty in the
parameters gives, in general, asymmetric error bars, which can be an additional desirable features in
situation in which the uncertainties make parameters that should be positive, compatible with zero
or negative values. Apart from the above considerations, the fitted parameters appear compatible
with what is expected for this source. It is however important to consider in more detail the statistical
significance of these results by applying some \emph{goodness of fit} test, which we will discuss in the
following section.

\subsection{\label{sec:sta}Statistical significance}

After obtaining the fit parameters, we can now proceed with the last step, i.e. discuss their statistical
significance. To this end we will consider, for each of the nine fitted SEDs, the residuals of the fit,
i.e. the differences between the observed points and the value of the fitted SED at the frequency of
the observed points. We then apply the KS test to verify if the residuals are normally
distributed (the $\mathscr{N} _{0} ''$ null-hypothesis of page~\pageref{text:nulhyp002}).
Code for the calculation of the relevant statistics $D _{N}$ (cf. Eq.~(\ref{eq:kolsmista})) is available,
for instance in Ref.~\refcite{bib:1992nricPress}. In this study, however, we used the functions that are
included in Mathematica\textregistered{} since version $8.0$: the reason for this is the fact that these
functions already implement a method for the Monte Carlo approach that we briefly mentioned at the very
end of Sec.~\ref{sec:stasig} on pag.~\pageref{text:MonCar}. A first application of the KS test at the
$5\%$ confidence level, shows that the residuals are \emph{not} normally distributed, i.e. the KS
test fails. Following this result, we applied the KS test again, this time at the $10\%$ confidence level:
again the test failed in all cases, showing also at this confidence level that the residuals are not normally distributed

Failure of the KS test shows that the statistical significance of the fits should be carefully re-considered. In this
case it might be actually possible to explain the reason for this failure\footnote{To
have a more clear understanding of the situation, in each case we then decided to divide the data in two groups,
data at low energy (i.e. within the Synchrotron region of the spectrum) and data at very high energy
(i.e. within the Compton region of the spectrum). For each dataset we then calculated the residuals for the
low-energy subset of the data and for the high energy subset of the data. We then applied to all these subsets
the KS test (we called this procedure \emph{piecewise KS test}, as for each dataset the KS test for ${\mathscr{N}} _{0} ''$
is applied separately to low- and high-energy data). Surprisingly enough, in all these cases the KS test confirmed
normality of the residuals. Further discussion of this point can be found elsewhere\cite{bib:2011ApJ73314Mankuzhiyil}.
For our present purpose this further analysis is not necessary and we will ignore it here.}, but we will not proceed
here in this direction. We will instead draw a bold conclusion and emphasize the fact that, in absence of other
explanations, the failure of the KS test could already bring us to the conclusion that the model might require
improvements: existing observations, even in presence of a successful fit with reasonable values for the parameters
and for the reduced $\chi ^{2}$, could be enough to show the inadequacy of the model. We could have never reached
this result, had we not proceeded through a rigorous statistical approach and we recognize, in this way, the extreme
importance of an in depth statistical analysis to put to the best use our models and the related observations.

\section{Conclusion}

In this contribution we have discussed all the steps that are required to perform a rigorous statistical analysis
on simultaneous multi-wavelength datasets of blazars. Although it is challenging to obtain good datasets, because
observations are often made difficult by the necessity to have several instruments simultaneously
available in absence of observational constraints, we find really exciting to think that such observations (which
probe several order of magnitudes in the source spectrum with different instruments and techniques) could be already
at a good enough level to allow us to discriminate between different emission models. The importance of a statistical
approach is, indeed, two-sided. On one side it can force us to improve our models, to make them compatible with the data. On
the other, it can help us to understand how to plan future instruments and observations and efficiently improve, where
it is most needed, the quality and amount of the data. We have finally shown, in the specific case of Mrk421, how this
analysis has been applied to a specific problem and how existing data could be already suggesting the need for refinements
of the emission model for this source.

\section*{Acknowledgments}

The author would like to acknowledge partial support from ICRANet.

\end{document}